\documentclass[trackchange]{aastex7}

\newcommand{\bfm}[1]{\mbox{\boldmath{$#1$}}}
\usepackage{float}
\usepackage{amsmath} 
\usepackage{bm}
\usepackage{placeins}
\usepackage{color}
\begin{document}

\title{Challenge in Arrokoth's single merger to achieve the shape's principal axis configuration}

\author[orcid=0009-0001-8584-1612]{Ketan Kamat}
%\altaffiliation{Georgia Institute of Technology}
\affiliation{Georgia Institute of Technology, Atlanta, Georgia 30332, United States}
\email[show]{kkamat6@gatech.edu}

\author[orcid=0000-0002-9840-2416]{Ryota Nakano}
%\altaffiliation{Georgia Institute of Technology}
\affiliation{Georgia Institute of Technology, Atlanta, Georgia 30332, United States}
\email{rnakano@gatech.edu}

\author[orcid=0000-0002-1821-5689]{Masatoshi Hirabayashi}
%\altaffiliation{Georgia Institute of Technology}
\affiliation{Georgia Institute of Technology, Atlanta, Georgia 30332, United States}
\email{thirabayashi@gatech.edu}

%\collaboration{all}{The Terra Mater collaboration}

\begin{abstract}

The cold-classical Kuiper Belt Object 486958 Arrokoth is a contact binary composed of two flattened lobes, Weeyo and Wenu, closely aligned along their principal axes, despite each lobe having a highly irregular shape. The object's smooth and relatively undamaged structure suggests the observed bilobate shape results from a gentle, low-velocity merger between the lobes. The existing hypotheses to explain such a merger include orbital energy dissipation from the protosolar nebula gas drag and Lidov-Kozai (LK) oscillations originating from an initially ultra-wide binary. However, what is missing is how mutual dynamics due to the lobes' shape irregularities impact their final orientations at the time of the soft merger. Here, we show that none of the proposed orbital evolution scenarios is sufficient to reproduce the contact along the lobes' longest principal axes. Implementing the full two-body problem method using finite element modeling, we numerically quantify the complex mutual interactions between Weeyo and Wenu, before the soft merger under the reported geophysical constraints and orbital configurations. All simulations demonstrate that the rotational states of both lobes become desynchronized shortly after their close approach, eventually leading to substantial misalignment along their principal axes. We also find that the lobes' mutual gravitational torque, destabilizing their aligned orientations, is several orders of magnitude higher than gas-driven torque, suggesting that gas drag plays a negligible role in stabilizing their orientations. The present study suggests the necessity of an additional process reconfiguring Arrokoth's shape after the merging process, possibly due to the Sky-forming impact.

\end{abstract}

%%\keywords{\uat{Galaxies}{573} --- \uat{Cosmology}{343} --- \uat{High Energy astrophysics}{739} --- \uat{Interstellar medium}{847} --- \uat{Stellar astronomy}{1583} --- \uat{Solar physics}{1476}}


\section{Introduction} 
486958 Arrokoth (2014 MU69) is a bilobate object located in the Kuiper Belt, first imaged by NASA’s New Horizons mission in 2019 \citep{Nelson2022Arrokoth,Benecchi2019HSTLightcurve}. As a cold classical Kuiper Belt Object (KBO), Arrokoth is considered to represent planetesimal formation within the protosolar nebula, a dense disk of gas and dust that filled the solar system during its earliest stages of formation \citep{Nesvorny2010KBBFormation,Johansen2007PlanetesimalFormation,Lorek2024FlattenedPlanetesimals}. The object is approximately 35 km long and consists of two irregularly shaped lobes of very similar composition (Wenu and Weeyo), joined by a narrow neck \citep{Protopapa2019UltimaThule,Porter2024Arrokoth,Amarante2020ArrokothDynamics}. The principal axes of Arrokoth’s lobes are aligned to approximately 5$^\circ$, and the smooth, undamaged overall structure suggests it originates from a soft merger \citep{Spencer2020ArrokothGeology,White2021Arrokoth}.

Arrokoth’s possible evolutionary pathway is that its two lobes gradually evolve from a co-orbiting binary into a merged configuration (\citealp{Stern2023ArrokothMounds}; \citealp{McKinnon2020Arrokoth}). This process likely facilitated a gentle, low-energy merger between Weeyo and Wenu, stabilizing their proximity interactions while preserving the overall shape of the final body \citep{Stern2021ArrokothNeck,Marohnic2020ArrokothMerger,Wandel2019UltimaThuleFormation}. While the exact pathway that leads to the soft merger is still under debate, multiple plausible mechanisms exist. 

The first proposed mechanism is that Arrokoth’s bilobate structure formed through the gradual orbital decay of a once co-orbiting binary of Weeyo and Wenu, likely driven by drag from the gas in the protosolar nebula \citep{McKinnon2020Arrokoth}. As the binary orbited within this nebular gas, it experienced drag forces that extracted angular momentum from the system. The loss of angular momentum caused the two bodies to spiral inward and eventually merge at very low velocities, preserving Arrokoth’s structural integrity and producing its currently observed shape.

Another proposed theory is a framework in which non-secular Lidov-Kozai (LK) cycles drive such collisions between Weeyo and Wenu over timescales comparable to the binary’s outer orbital period ($\sim$1000 yr) by inducing significant variations in eccentricity and inclination \citep{Grishin2020WideBinary}. Orbital evolution modeling suggests that these fluctuations can lead to a substantial reduction in the separation between the lobes, which ultimately facilitates their collision \citep{Grishin2020WideBinary}. Smoothed particle hydrodynamics (SPH) simulations further confirm that the impact speeds driven by the LK cycles could yield a gentle collision consistent with Arrokoth’s morphology \citep{Grishin2020WideBinary}. 

A more comprehensive model for Arrokoth’s formation is that the bilobate structure arose from a combination of dynamical effects \citep{Lyra2021}. In addition to gas drag from the protosolar nebula and LK oscillations, this theory incorporates the gravitational quadrupole moments ($J_2$) of the lobes and tidal friction between them. These combined effects were modeled within a framework termed KTJD (Kozai, tides, J2, and drag) and led to a feasible collision scenario where the lobes gradually spiraled inward and merged \citep{Lyra2021}.

The key question that remains unsolved, however, is how such a soft merger robustly builds Arrokoth's current unique shape, in which Weeyo and Wenu are aligned almost perfectly along their longest principal axes. Given their irregular shapes, mutual gravitational torque during close encounters likely disturbs their doubly synchronous state, which is a necessary condition for the soft merger along their longest principal axes \citep{Stern2019MU69}. As demonstrated in this study, none of the proposed formation theories above confirm reasonable conditions for Arrokoth to attain its current configuration via a one-time merger.

\section{Mutual dynamical characterization in the pre-merger phase}

This study examines how the hypothesized soft merger scenario may cause Weeyo and Wenu to tumble relative to each other during the pre-merger phase. We examine three orbital configurations to observe how the mutual attitude dynamics become unstable. Our primary focus is on the rotational state of each lobe, which is driven by the close encounter and becomes unstable on a much shorter timescale compared to the full merging process. Thus, our orbital settings only select small, key pieces of the proposed Arrokoth merging scenarios and mimic such configurations solely based on the mutual gravity interactions between the lobes. We employ 441 simulations, accounting for 49 shape combinations and three bulk densities across a variety of orbits. For the bulk density variation, we explore three cases: $235~\mathrm{kg\,m^{-3}}$, $367.5~\mathrm{kg\,m^{-3}}$, and $500~\mathrm{kg\,m^{-3}}$ \citep{Keane2022Arrokoth}. Further details are provided below.

For each orbital configuration, we produce seven shape variants for both Wenu and Weeyo, leading to 49 binary shape combinations in total (Tables \ref{tab:wenu_shapes} and \ref{tab:weeyo_shapes}). Shape 0 represents the baseline geometry for each lobe, while shapes 1-6 introduce symmetric perturbations of $\pm$250~m to the major ($a$), intermediate ($b$), and minor ($c$) semi-axes. Wenu and Weeyo have volumes of $1.98 \times 10^{11}$~\text{m}$^3$ and $1.10 \times 10^{11}$~\text{m}$^3$, respectively, based on shape models of Arrokoth's lobes \citep{Keane2022Arrokoth}. To ensure that the sensitivity analysis isolates geometric effects without altering the system's mass, the volume of each generated shape is kept constant. The $\pm$250~m perturbation applied to the major and intermediate semi-axes correspond to the maximum observational uncertainties reported by the New Horizons team \citep{Spencer2020ArrokothGeology}. These 14 individual shape models (seven models for each) serve as the basis for the 49 binary shape combinations in simulated orbits. For each shape combination, we track the total in-plane prolateness ($\chi_p$) using Equation ({\ref{eq:in_plane}}):
\begin{eqnarray}
\chi_p = \sum_{i=1}^{2} \frac{a_i}{b_i}
\label{eq:in_plane}
\end{eqnarray}
where subscript $i \: (= 1,2)$ represents the quantity for Wenu and Weeyo, respectively.

\begin{table}[!b]
\centering
\caption{{Dimensions and axial ratios of the unique Wenu shapes.}}
\label{tab:wenu_shapes}
\begin{tabular}{ccccccc}
\hline
Shape & $a$ [m] & $b$ [m] & $c$ [m] & $a/c$ & $b/c$ & $a/b$ \\ \hline
0 & 10600 & 9950 & 4525 & 2.34 & 2.20 & 1.07 \\
1 & 10350 & 9950 & 4634.30 & 2.23 & 2.15 & 1.04 \\
2 & 10850 & 9950 & 4420.74 & 2.45 & 2.25 & 1.09 \\
3 & 10600 & 10200 & 4414.09 & 2.40 & 2.31 & 1.04 \\
4 & 10600 & 9700 & 4641.62 & 2.28 & 2.09 & 1.09 \\
5 & 10318.78 & 9686.03 & 4775 & 2.16 & 2.03 & 1.07 \\
6 & 10905.54 & 10236.80 & 4275 & 2.55 & 2.39 & 1.07 \\ \hline
\end{tabular}
\end{table}

\begin{table}[!t]
\centering
\caption{Dimensions and axial ratios of the unique Weeyo shapes.}
\label{tab:weeyo_shapes}
\begin{tabular}{ccccccc}
\hline
Shape & $a$ [m] & $b$ [m] & $c$ [m] & $a/c$ & $b/c$ & $a/b$ \\ \hline
0 & 7875 & 6925 & 4875 & 1.62 & 1.42 & 1.14 \\
1 & 7625 & 6925 & 5034.84 & 1.51 & 1.38 & 1.10 \\
2 & 8125 & 6925 & 4725 & 1.72 & 1.47 & 1.17 \\
3 & 7875 & 7175 & 4705.14 & 1.67 & 1.52 & 1.10 \\
4 & 7875 & 6675 & 5057.58 & 1.56 & 1.32 & 1.18 \\
5 & 7680.53 & 6753.99 & 5125 & 1.50 & 1.32 & 1.14 \\
6 & 8085.04 & 7109.70 & 4625 & 1.75 & 1.54 & 1.14 \\ \hline
\end{tabular}
\end{table}

The first two orbital configurations are designed to represent merger scenarios driven by the LK effect. The strength of the LK effect strongly depends on the semi-major axis of the binary orbit \citep{Grishin2020WideBinary,Lyra2021}. The first set of orbits has a semi-major axis that exceeds a critical threshold for triggering non-secular LK collisions. In this regime, the LK perturbation operates at full strength, inducing chaotic oscillations in both eccentricity and inclination. This orbital configuration is termed as ``Non-Secular" because such non-secular LK collisions cause Arrokoth's gentle merger \citep{Grishin2020WideBinary}. The subsequent orbital configuration, later labeled ``Precession", consists of orbits that fall within a domain where the LK effect is largely suppressed by precessional effects of the binary orbit \citep{Grishin2020WideBinary}. These simulations aim to demonstrate that even in a weak LK effect regime, Weeyo and Wenu still exhibit significant mutual rotational instability.

The third orbital configuration, later termed ``Min-Separation", is designed to represent the final stage of orbital decay driven by gas drag in the protosolar nebula, which occurs on the order of millions of years \citep{McKinnon2020Arrokoth}. For these simulations, Weeyo and Wenu are initialized in circular orbits just under the critical stability separation. Since the critical stability separation is a function of the shape combination of the lobes, there are 49 distinct minimum-separation orbits. Below this critical stability separation, the lobes cannot maintain energetically stable relative equilibria \citep{scheeres2009stability}.

\begin{deluxetable*}{lccccccc}[ht!]
\tablecaption{Orbital configurations, assumed bulk densities, semi-major axis ($a$), eccentricity ($e$), periapsis distance ($r_p$), apoapsis distance ($r_a$), total integration time, and orbital period ($T$) for each merger scenario.}
\label{tab:orbit_data}
\tablewidth{0pt}
\tabletypesize{\small}
\tablehead{
Orbital Configuration & Bulk Density & $a$ & $e$ & $r_p$ & $r_a$ & Integration Time & $T$ \\
 & [kg m$^{-3}$] & [km] & [-] & [km] & [km] & [yr] & [yr]
}
\startdata
Non-Secular & 235 & 6442.34 & 0.99 & 50 & 12834.68 & 45 & 14.81 \\
Non-Secular & 367.5 & 7209.36 & 0.99 & 50 & 14368.72 & 45 & 14.02 \\
Non-Secular & 500 & 7818.57 & 0.99 & 50 & 15587.15 & 45 & 13.57 \\ 
\hline
Precession & 235 & 1651.88 & 0.97 & 50 & 3253.76 & 8 & 1.92 \\
Precession & 367.5 & 1917.38 & 0.97 & 50 & 3784.77 & 8 & 1.92 \\
Precession & 500 & 2124.61 & 0.98 & 50 & 4199.23 & 8 & 1.92 \\ \hline
Min-Separation & 235 & 23.08--24.57 & 0 & 23.08--24.57 & 23.08--24.57 & 0.0051--0.0102 & 0.0031--0.0033 \\
Min-Separation & 367.5 & 23.08--24.57 & 0 & 23.08--24.57 & 23.08--24.57 & 0.0036--0.0065 & 0.0024--0.0027 \\
Min-Separation & 500 & 23.08--24.57 & 0 & 23.08--24.57 & 23.08--24.57 & 0.0014--0.0035 & 0.0021--0.0023 \\
\enddata
\end{deluxetable*}

We employ the full two-body problem finite element model (F2BPFEM) to determine the mutual dynamical states of Weeyo and Wenu in the pre-merger phase(\citealp{Yu2019FEMFullTwoBody}; \citealp{nakano2022nasa}). Using this model, we inherently account for the irregular shapes of Weeyo and Wenu by capturing the mutual gravitational potential between them. Key irregular shape terms, such as those resulting from shape elongation, lead to more complex interactions than seen in a point-mass approximation model which requires explicit $J_2$ quadrupole terms. The resulting mutual gravitational torque, driven by the lobes' shape irregularities, serves as the primary driver for their rotational instability. The LK effect is not explicitly modeled, as our focus is on merger feasibility based on the rotational behavior of the lobes.

\begin{figure}[H]
\centering
\includegraphics[width=0.5\linewidth]{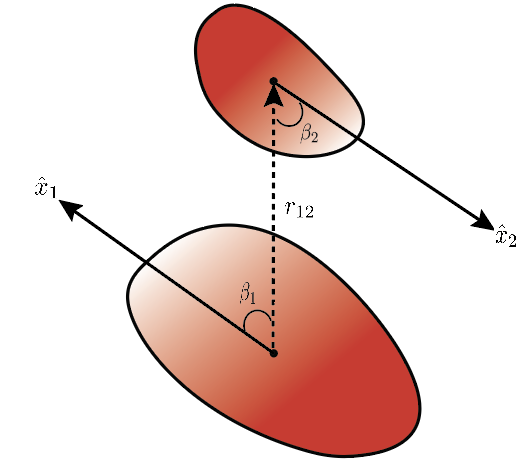}
\caption{Schematic definition of the collision angle ($\beta$). The total collision angle $\beta$ is calculated as the sum of the angular misalignments ($\beta_1$ and $\beta_2$) of each lobe's longest principal axis ($\hat{x}_1$ and $\hat{x}_2$) relative to the mutual line of centers ($r_{12}$).}
\label{fig:beta_angle}
\end{figure}

To analyze the system's rotational evolution, we track the orientation of Weeyo's body-fixed frame relative to Wenu's. For all orbital configurations, we characterize out-of-plane departures by monitoring the roll and pitch components of the 1-2-3 Euler angle sequence {\citep{agrusa2021excited}}. In our convention for each lobe’s body-fixed frame, the $\hat {\bfm x}$-axis aligns with the minimum moment of inertia (corresponding to the longest axis), the $\hat {\bfm y}$-axis with the intermediate moment of inertia, and the $\hat {\bfm z}$-axis with the maximum moment of inertia (corresponding to the shortest axis). The same definition applies to the subscripts to these quantities. To better characterize the misalignment of the lobes' longest principal axes upon a theoretical merger, we adopt a collision angle ($\beta$) in place of the standard yaw angle ({Figure}~\ref{fig:beta_angle}). 

The collision angle ranges from $0^{\circ}$ to $180^{\circ}$ and is calculated by summating the magnitude of angular misalignment for both lobes' long axes relative to the line of centers. In this convention, $\beta = 0^{\circ}$ corresponds to a state of alignment in which the long axes of both lobes point directly along the line of centers ($\bfm r_{12}$), defined as the position vector of Weeyo's center of mass relative to Wenu's. Equation (\ref{coll_angle}) provides the mathematical definition of this quantity:

\begin{eqnarray}
\beta = {\sum_{i=1}^{2} \beta_i} = \cos^{-1} \left ( \frac{|\hat{\bfm x}_1 \cdot \bfm r_{12}|}{\vert \bfm r_{12} \vert} \right) + \cos^{-1} \left ( \frac{|\hat{\bfm x}_2 \cdot \bfm r_{12}|}{\vert \bfm r_{12} \vert} \right)
\label{coll_angle}
\end{eqnarray}
where $\hat{\bfm x}_1$ and $\hat{\bfm x}_2$ are the unit vectors toward the lobes' long axes. We also compute the short-axis spin ratio ($\sigma_i$), defined in Equation (\ref{short_axis_spin}) as the ratio of angular momentum about the $\bfm z$-axis of each lobe to its total body angular momentum. At the start of each simulation, both Weeyo and Wenu are initialized with pure short-axis spin ($\sigma_i = 1$):
\begin{eqnarray}
\sigma_i = \frac{[\bfm I]_i \bfm \omega_i \cdot \hat {\bfm z}_i}{\vert [\bfm I]_i \bfm \omega_i \vert}
\label{short_axis_spin}
\end{eqnarray}
where $ [\bfm I]_i$ is the moment of inertia tensor for either lobe and $\bfm \omega_i$ represents the angular velocity vector for a given lobe.

Across all three orbital configurations, Weeyo and Wenu are initialized at apoapsis with their principal axes perfectly aligned. We employ the state of relative equilibrium to achieve this condition \citep{scheeres2009stability}. In our problem, the equation by \cite{scheeres2009stability} can be written as follows: 
\begin{eqnarray}
    \dot{\theta}^2 = \frac{\mathcal{G}(M_1 + M_2)}{{\vert \bfm r_{12} \vert}^3} \left( 1 + \frac{3 C_{12}}{2{\vert  \bfm r_{12} \vert}^2}  \right)
\label{eq:equilibrium_spin}
\end{eqnarray}
where $\mathcal{G}$ is the gravitational constant, $M_1$ and $M_2$ are the masses of Wenu and Weeyo, and $C_{12}$ is given as follows:
\begin{eqnarray}
C_{12} = \sum_{i=1}^2 \left \{ Tr ([\bfm I]_i) - 3 I_{x, i} \right \} \label{Eq:C12}
\end{eqnarray}
In Equation (\ref{Eq:C12}), $Tr$ is the trace operator, and $ I_{x,i}$ is the moment of inertia component along the long axis of a given lobe.

While we apply this initial orbit setting to all configurations, it may not perfectly represent the highly elliptic cases, i.e., the non-secular and precession orbits, where the orbital angular velocity varies significantly. These elliptic configurations result from LK resonances, which drive an initially low-eccentricity orbit to become highly eccentric \citep{Grishin2020WideBinary}. To model the rotational response during this sudden transition to high eccentricity, we initialize the lobes in a doubly synchronous state within a circular orbit prior to the appearance of the high-eccentricity state, later known as the pre-phase circular orbit. Since variations in orbital angular velocity generally disrupt synchronicity, we consider this initial setting to represent a conservative case for the system's stability.

\section{Immediate appearance of rotational instability}

We model each lobe as a tetrahedral mesh composed of 362 vertices and 1388 tetrahedrons to ensure high-fidelity gravitational resolution. The dynamics are propagated using a Runge-Kutta-Fehlberg 4(5) (RKF45) integration scheme with a variable step size ranging between 500 and 1,000 s \citep{Montenbruck2000}. We monitor the system's total energy throughout each simulation, maintaining a strict conservation range where the system energy variation remains on the order of $10^{-6}\text{--}10^{-5}$\% across all orbits (Table \ref{tab:energy_variation}). These results demonstrate that the observed desynchronization and tumbling are legitimate dynamical behaviors of the binary system rather than integrator performance or step-size sensitivity.

\begin{table}[h]
\centering
\caption{Mean System Energy Variation for each type of orbit across all simulations.}
\begin{tabular}{lcc}
\hline
\hline
Orbital Configuration & Mean System Energy Variation (\%) \\
\hline
Non-Secular & $1.82 \times 10^{-5}$ \\
Precession & $9.39 \times 10^{-6}$ \\
Min-Separation & $6.67 \times 10^{-6}$ \\
\hline
\end{tabular}
\label{tab:energy_variation}
\end{table}

\begin{figure}[H]
\centering
\includegraphics[width=0.8\linewidth]{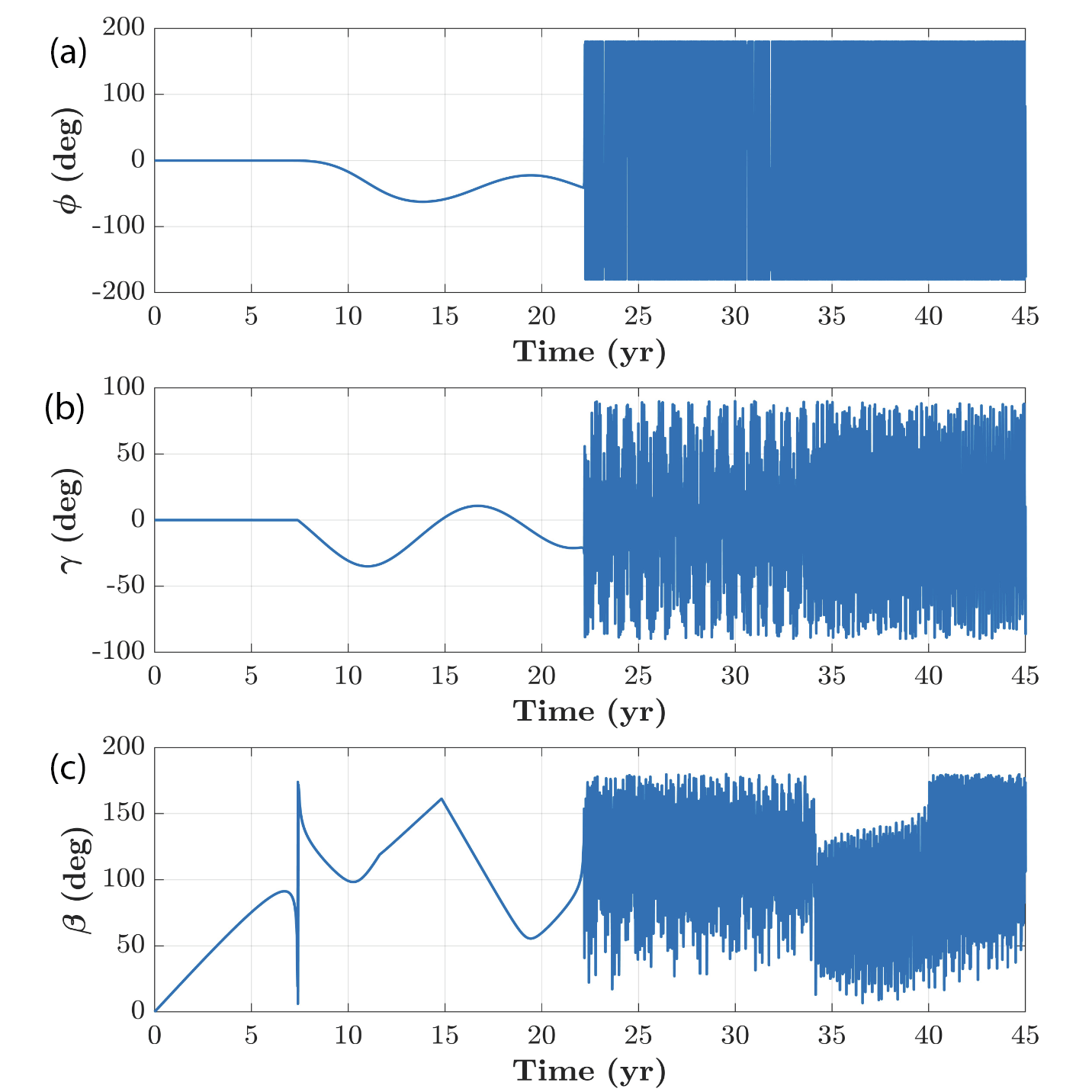}
\caption{Time evolution of the relative rotational modes of Weeyo with respect to Wenu for a non-secular orbit, assuming a bulk density of 235~kg m$^{-3}$ and default lobe shapes. Panel (a) shows the roll angle $\phi$, panel (b) shows the pitch angle $\gamma$, and panel (c) shows the collision angle $\beta$.}
\label{fig:ypr_plot_non_sec}
\end{figure}

For the non-secular orbits, the rotational states display an immediate departure from the starting aligned configuration. Figure \ref{fig:ypr_plot_non_sec} illustrates this behavior for the baseline case (Shape 0, 235~kg~m$^{-3}$), where the collision angle drifts steadily before rapidly climbing toward 180$^\circ$ as the lobes approach their first periapsis encounter ($\sim$7.5 yr). While the roll and pitch angles initially maintain relative stability, they begin to display significant oscillatory behavior after this first close encounter with amplitudes reaching peaks of approximately 60$^\circ$ and 35$^\circ$ respectively  between $\sim$7.5 and $\sim$22~yr. Upon the second periapsis at $\sim$22~yr, the relative orientation becomes increasingly erratic and shows no sign of recovery, as the subsequent proximity encounter ($\sim$35~yr) further amplifies the rotational instability. This instability develops on a timescale significantly shorter than the long-term evolution driven by the LK effect, which induces orbital oscillations leading to a merger over several thousand years.

\begin{figure}[H]
\centering
\includegraphics[width=0.8\linewidth]{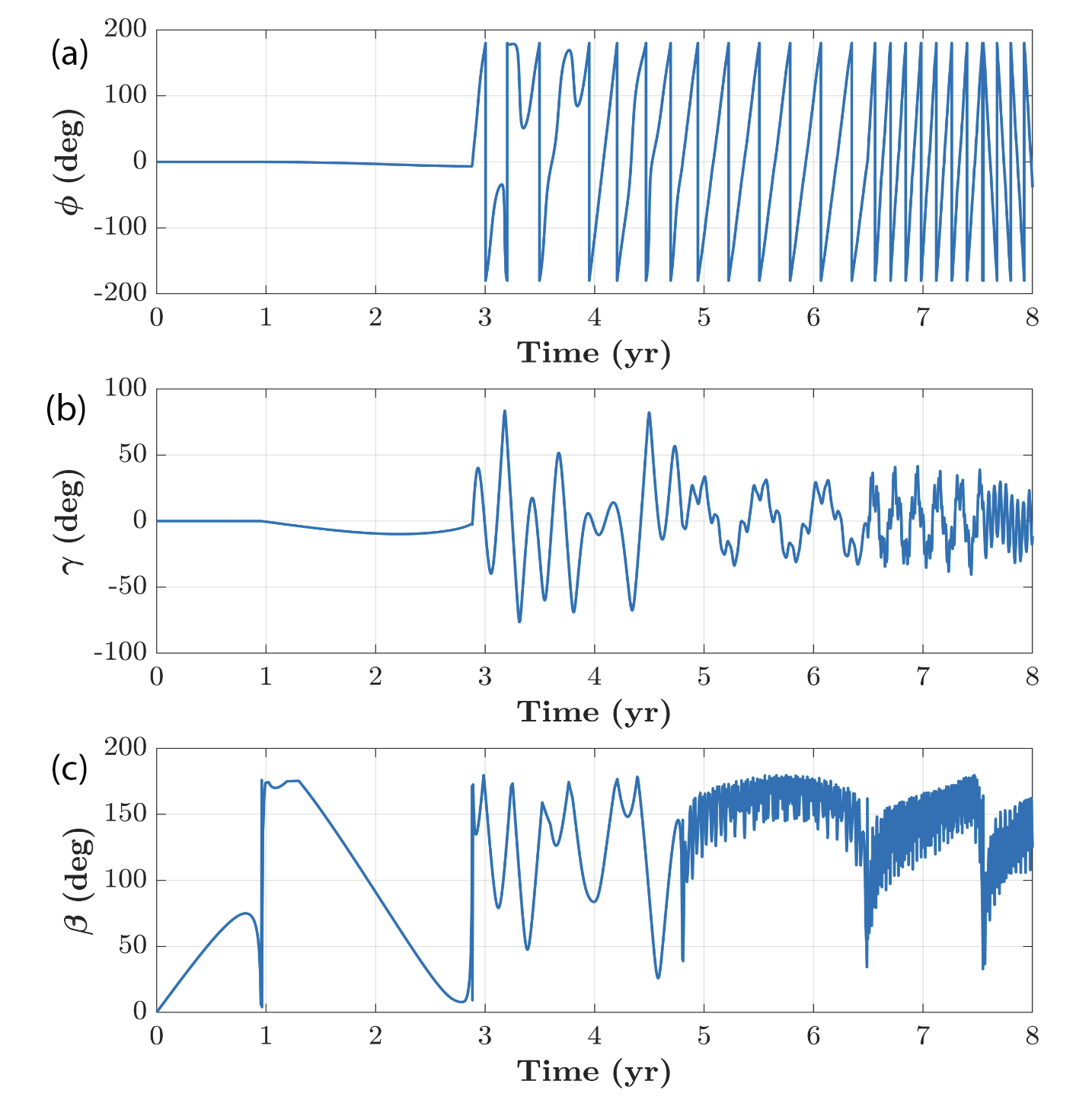}
\caption{Time evolution of the relative rotational modes of Weeyo with respect to Wenu for a precession orbit, assuming a bulk density of 235~\text{kg m}$^{-3}$ and default lobe shapes. Panel (a) shows the roll angle $\phi$, panel (b) shows the pitch angle $\gamma$, and panel (c) shows the collision angle $\beta$.}
\label{fig:ypr_plot_precession}
\end{figure}

The precession orbits similarly result in widespread rotational instability. As shown in Figure~\ref{fig:ypr_plot_precession} for the baseline shape and density settings, the initial evolution of the collision angle is similar to the non-secular orbit, exhibiting a steady linear drift from the start of the simulation. This misalignment climbs to a local peak of approximately 75$^\circ$ before swiftly transitioning toward 180$^\circ$ during the first periapsis encounter at $\sim$1~yr. While the roll and pitch angles remain relatively stable following this initial encounter, significant misalignments occur across all three rotational modes after the second periapsis at $\sim$3~yr. Most notably, the roll mode initiates full $\pm$180$^\circ$ oscillations, and the collision angle again reaches a maximum misalignment near 180$^\circ$. Following the third periapsis encounter at $\sim$5~yr, the system transitions into a sustained chaotic regime; in this state, the collision angle oscillates rapidly between 45$^\circ$ and 180$^\circ$, the pitch mode oscillates with amplitudes of 40$^\circ$, and the roll mode continues its full-scale oscillations of $\pm$180$^\circ$. As observed in the non-secular orbits, Weeyo and Wenu do not return to a doubly synchronous state.

\begin{figure}[H]
\centering
\includegraphics[width=0.8\linewidth]{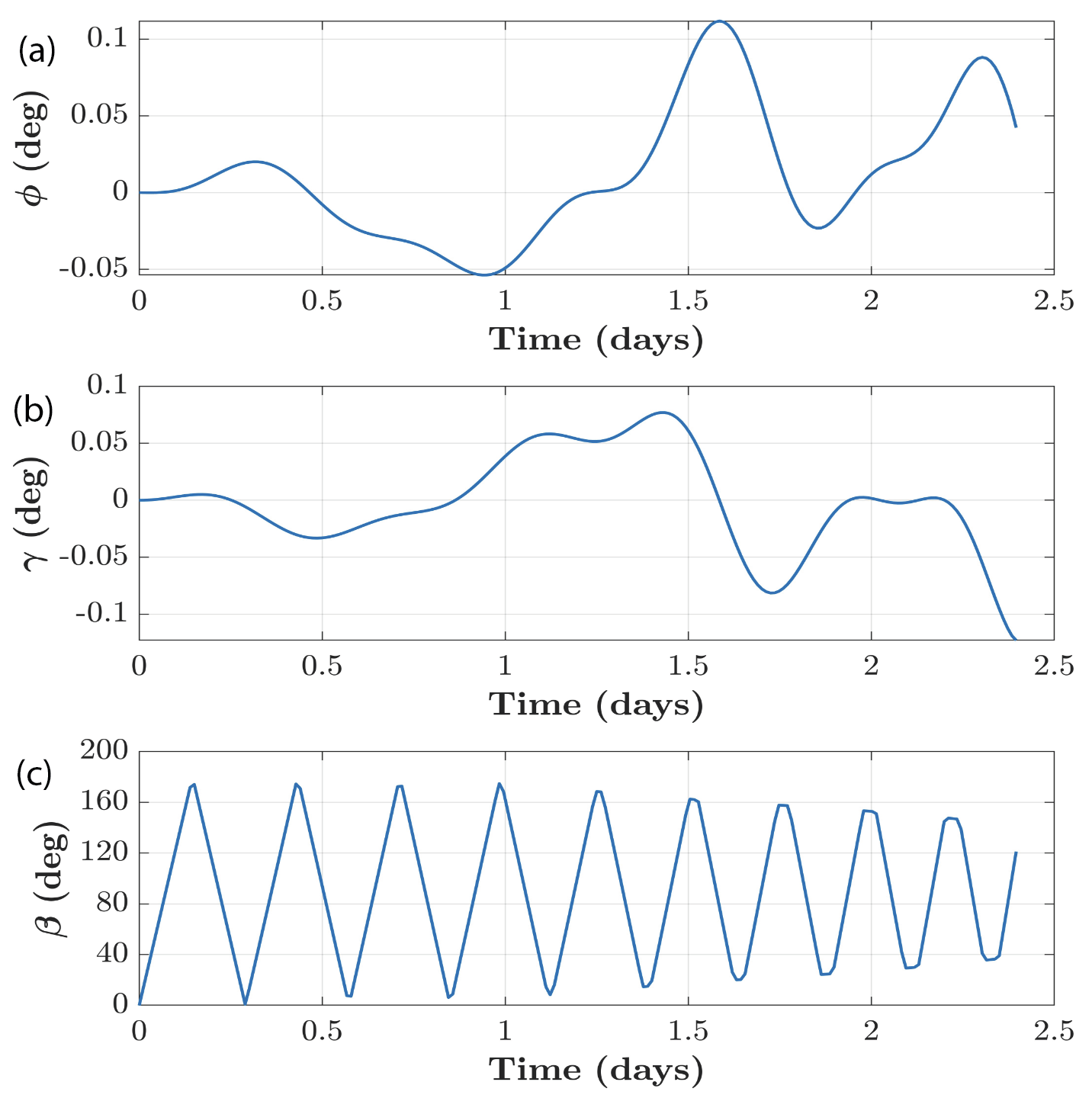}
\caption{Time evolution of the relative rotational modes of Weeyo with respect to Wenu for a min-separation orbit, assuming a bulk density of 235~\text{kg m}$^{-3}$ and default lobe shapes. Panel (a) shows the roll angle $\phi$, panel (b) shows the pitch angle $\gamma$, and panel (c) shows the collision angle $\beta$.}
\label{fig:ypr_plot_ms}
\end{figure}

Finally, the minimum-separation orbits exhibit distinct dynamical modes characterized by rapid desynchronization. This accelerated instability occurs due to the extremely close proximity of the lobes and resulting strong mutual gravitational torque. Unlike the other configurations, this setup leads to a collision between Weeyo and Wenu within 0.5--4 days. Across these orbits, the roll and pitch modes remain near-zero, but the collision angle deviates from the start of the simulation. For the default case illustrated in Figure~\ref{fig:ypr_plot_ms}, the lobes collide within $\sim$2.4 days. Specifically, the collision angle exhibits a periodic evolution, oscillating between $0^\circ$ and $\sim$$180^\circ$ roughly every 0.25 days as the lobes desynchronize. As the system approaches impact, the amplitude of the collision angle oscillation gradually reduces, resulting in a value of $\sim$$120^\circ$ at the moment of collision. For these orbits, increasing the bulk density of the lobes results in heightened instability and accelerated collision times.

\section{Infeasibility of a Single Merger and Proposed Post-Merger Realignment Mechanism}

Our mutual dynamics simulations showed that none of the orbital configurations could yield an aligned configuration between Weeyo and Wenu in the pre-merger phase. Both lobes quickly desynchronized from their initially doubly synchronous state and experienced rotational instability after a few orbital periods. 

\begin{figure}[th!]
\centering
\includegraphics[width=0.4\linewidth]{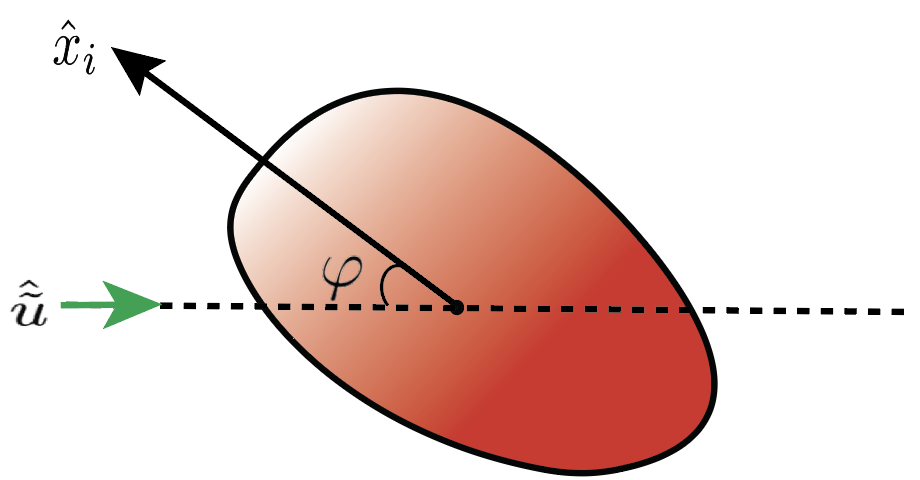}
\caption{Geometry of the aerodynamic incidence angle ($\varphi$) for a given lobe. $\varphi$ represents the orientation of a lobe's long axis ($\hat{x}_i$) relative to the incident velocity vector of the surrounding gas ($\hat{\tilde{u}}$).}
\label{fig:phi_angle}
\end{figure}

One possibility mitigating the unstable rotational state is the continuous exposure of the lobes to gas drag within the protosolar nebula during the gradual orbital decay \citep{McKinnon2020Arrokoth, Lyra2021}. To assess whether the protosolar nebula gas can stabilize the proximity dynamics between Arrokoth's lobes, we use an aerodynamic model for ellipsoids with fluid parameters for this environment \citep{Zastawny2012DragLiftTorque}. To calculate the pitching torque on the ellipsoid, we first calculate the pitching torque coefficient ($C_T$). As shown in Equation (\ref{eq:c_t}), this is dependent upon the ellipsoid's Reynolds number ($Re$) and angle of incidence ($\varphi$) relative to the fluid (Figure~\ref{fig:phi_angle}). This model uses ten fitting parameters derived from direct numerical simulation results for a specific ellipsoid shape. The $C_T$ quantity is obtained as follows:
\begin{equation}
C_T = \left( \frac{0.935}{Re^{0.146}} - \frac{0.469}{Re^{0.145}} \right) \sin^{\alpha_1} \varphi \cos^{\alpha_2} \varphi
\label{eq:c_t}
\end{equation}
where 
\begin{eqnarray*}
    \alpha_1 &=& 0.116 + 0.748 Re^{0.041} \\
    \alpha_2 &=& 0.221 + 0.657 Re^{0.044}
\end{eqnarray*}

 $C_T$ is strictly non-negative because the incidence angle $\varphi$ is constrained to the first quadrant, ensuring positive trigonometric products, while the bracketed term $\left( 0.935/Re^{0.146} - 0.469/Re^{0.145} \right)$ remains positive for all $Re \in [1, 300]$.
Using Equation (\ref{eq:T_p}) and the dynamic pressure yields the pitching torque ($\tau_P$) value:
\begin{eqnarray}
\tau_P =  \frac{1}{2} C_T \rho_a \tilde{u}^2 \frac{\pi}{8} d_p^3
\label{eq:T_p}
\end{eqnarray}
where $\tilde{u}$ is the surrounding gas speed, and $\rho_a$ is the gas density, and $d_p$ is the effective diameter.

Within the protosolar nebula, the $\rho_a$ value is estimated to be $3 \times 10^{-11}$ kg\,m$^{-3}$ and a fixed headwind velocity of 50 m s$^{-1}$ \citep{Lyra2021,McKinnon2020Arrokoth} is assumed. Under these conditions, the corresponding $Re$ is in the range of 10–100, calculated using embedded sphere models based on the equivalent radii of Weeyo and Wenu \citep{Zastawny2012DragLiftTorque}. The referenced aerodynamic model calculates the gas torque acting on each lobe solely based on flow incident along its long axis, producing only a pitching torque \citep{Zastawny2012DragLiftTorque}. To approximate the full effect of the gas torque ($\tau_{gas}$), we assume that all the components are of similar magnitudes and thus multiply the pitching torque by $\sqrt{3}$ to compute the full scale torque. If the resulting total gas torque is comparable to or greater than the mutual gravitational torque ($\tau_{grav}$), it may be sufficient to mitigate rotational instability.

For all the cases considered in this study, the gas torque upon the lobes is several magnitudes lower than their mutual gravitational torque for the majority of the orbit. The ratio of gas to gravitational torque reaches its minimum near periapses, where the lobes are in proximity, and mutual gravitational interactions dominate. When the lobes are the farthest apart, there are brief periods where the induced gas torque exceeds the gravitational torque. This occurs due to the weakening of mutual interactions between Weeyo and Wenu at larger separations. However, these instances are short-lived and insufficient to restore the system to a doubly synchronous state. 

\begin{figure}[H]
\centering
\includegraphics[width=0.8\linewidth]{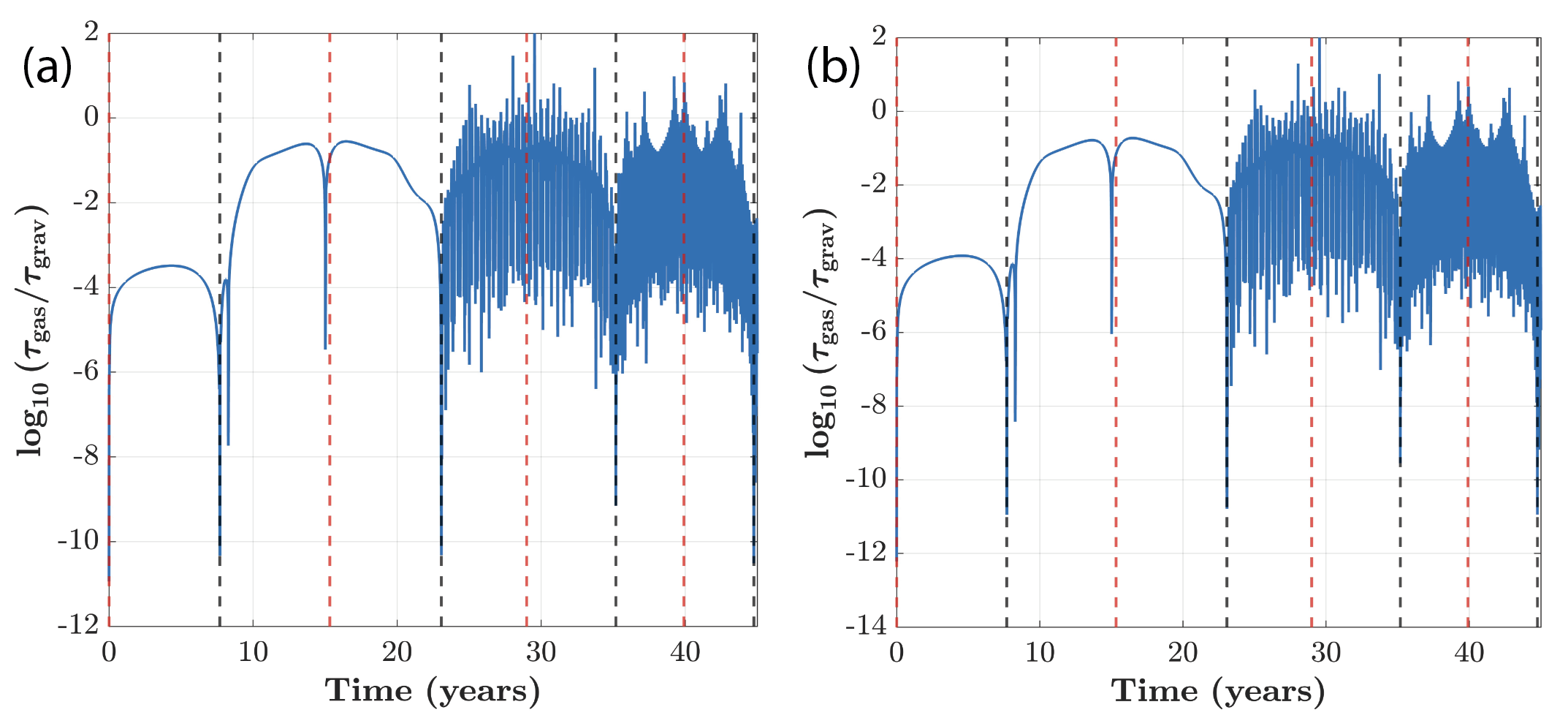}
\caption{Ratio of gas torque to gravitational torque on Weeyo for $Re=10$ in panel (a) and $Re=100$ in panel (b). Apoapsis times are shown in red and periapsis times are shown in black for a non-secular orbit with the 235 kg m$^{-3}$ bulk density and default lobe shapes.}
\label{fig:pitch_grav_ratio}
\end{figure}

Figure~\ref{fig:pitch_grav_ratio} shows a non-secular orbit, in which the gas torque on Weeyo is $\sim${1.36}$\%$ and $\sim${0.89} $\%$ of the gravitational torque on average for $Re = 10$ and $Re = 100$, respectively. On very limited occasions (i.e., at apoapses), the gas torque exceeds the gravitational torque, but this effect is overall negligible. The precession orbits exhibit similar behavior to the non-secular orbits, with gas torque remaining consistently weaker than mutual gravitational torque. For the minimum-separation orbits, the gas-to-gravitational torque ratio remains extremely low up until collision. In contrast to other cases, the minimum-separation orbits show that Weeyo and Wenu maintain nearly pure short-axis rotation, but this does not lead to a different consequence. The gravitational torque is always dominant; therefore, the gas torque is never able to stabilize the rotational state of each lobe. 

\begin{figure}[H]
\centering
\includegraphics[width=0.75\linewidth]{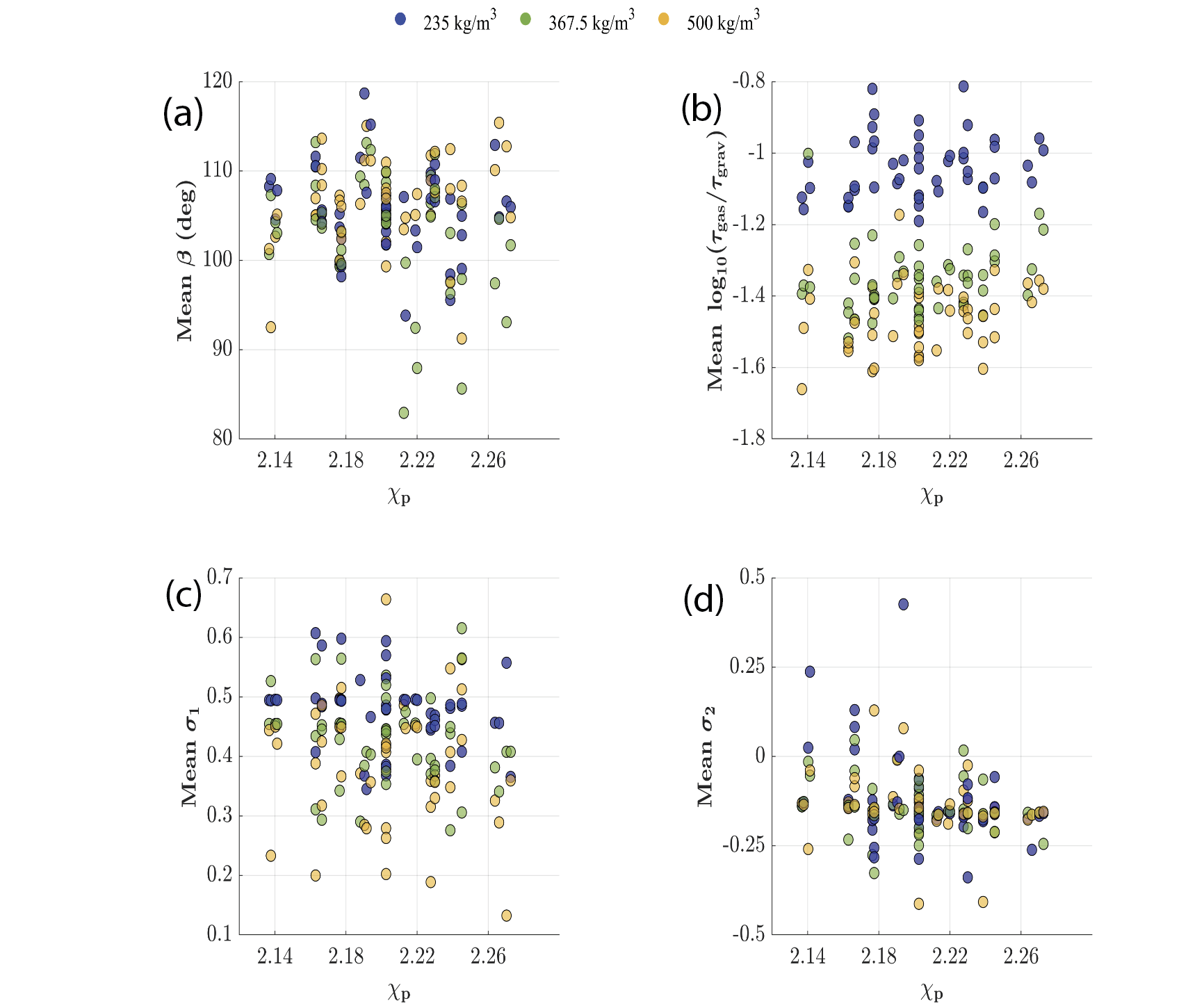}
\caption{Statistical distribution for the non-secular cases: (a) Mean collision angle, (b) Mean ratio of gas torque to gravitational torque on Weeyo for $Re=10$, (c) Mean Wenu short-axis spin ratio, and (d) Mean Weeyo short-axis spin ratio.}
\label{fig:NS_summary_plot}
\end{figure}

\begin{figure}[H]
\centering
\includegraphics[width=0.75\linewidth]{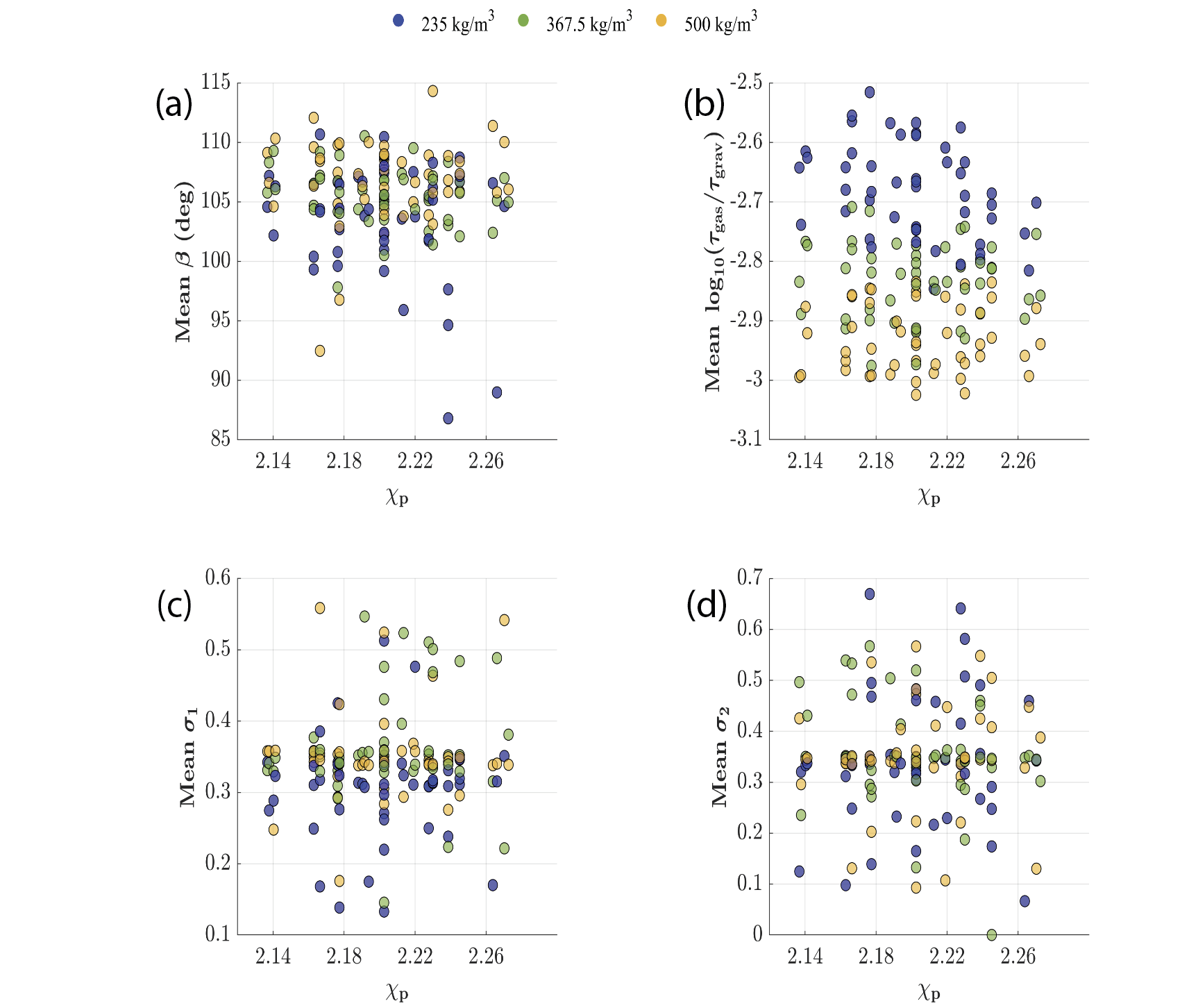}
\caption{Statistical distribution for the precession cases: (a) Mean collision angle, (b) Mean ratio of gas torque to gravitational torque on Weeyo for $Re=10$, (c) Mean Wenu short-axis spin ratio, and (d) Mean Weeyo short-axis spin ratio.}
\label{fig:PR_summary_plot}
\end{figure}

\begin{figure}[H]
\centering
\includegraphics[width=0.75\linewidth]{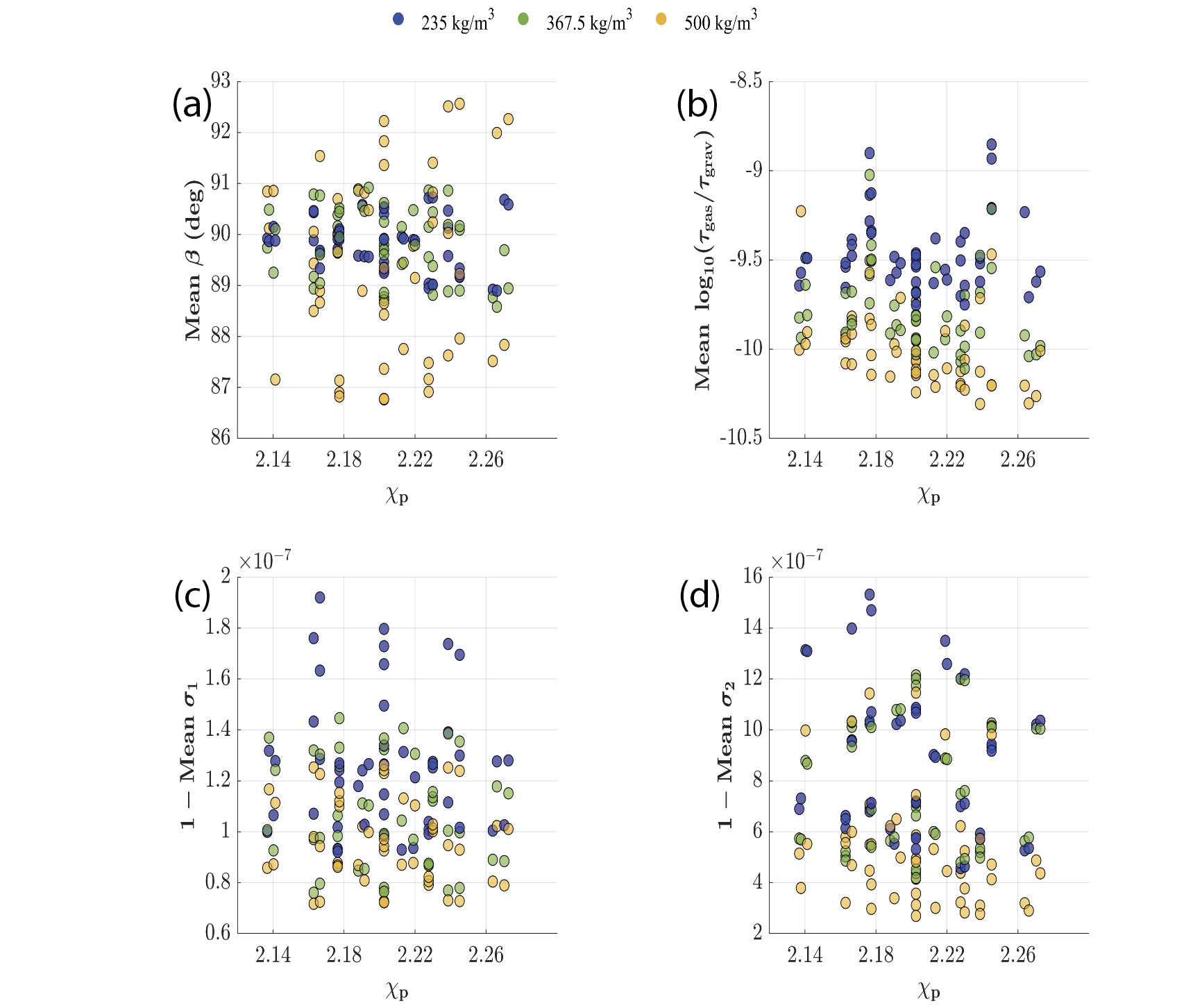}
\caption{Statistical distribution for the min-separation cases: (a) Mean collision angle, (b) Mean ratio of gas torque to gravitational torque on Weeyo for $Re=10$, (c) Deviation of mean Wenu short-axis spin ratio from unity, and (d) Deviation of mean Weeyo short-axis spin ratio from unity.}
\label{fig:MS_summary_plot}
\end{figure}

Based on the above analysis of all simulated orbital configurations, the gas torque is consistently found to be significantly weaker than the mutual gravitational torque. These results demonstrate that protosolar nebula gas is not a viable mechanism for stabilizing the mutual dynamics between Weeyo and Wenu and facilitating a single soft merger. To better characterize the dynamic evolution of Weeyo and Wenu, we examine key metrics across all simulated orbits. Figures~\ref{fig:NS_summary_plot}, \ref{fig:PR_summary_plot}, and \ref{fig:MS_summary_plot} present a statistical summary of the rotational evolution across all simulated cases, grouped by orbital configuration. These figures track the mean collision angle, gas-to-gravitational torque ratio upon Weeyo, and the short-axis spin ratios of both lobes. Collectively, the following results demonstrate that the mutual instability observed between Weeyo and Wenu across all simulated orbits makes a single soft merger scenario for Arrokoth highly infeasible.

Analysis of the non-secular orbits reveals that the mean collision angles between Weeyo and Wenu are predominantly concentrated between $80^\circ$ and $120^\circ$ (Figure~\ref{fig:NS_summary_plot}a). There is no discernible relationship observed between the mean collision angle and either the assumed bulk densities or shapes of the lobes. The drag torque observed in this orbital configuration is the highest among the three, primarily because of the high degree of multi-axis rotation of the lobes. However, the gravitational torque remains the overwhelmingly dominant force, exerting a magnitude between $6$ to $50$ times greater than the gas torque (Figure~\ref{fig:NS_summary_plot}b). Furthermore, the non-secular orbits display a negative correlation between the gas-to-gravitational torque ratio and bulk density. This trend remains persistent across all simulated orbital configurations and occurs because the mutual gravitational torque scales directly with the mass of the lobes. In the non-secular regime, both Wenu and Weeyo depart from their pure short-axis spin modes and transition into multi-axis tumbling states. A majority of these cases produce a mean short-axis spin ratio for Wenu within a range of $0.3$ to $0.5$ (Figure~\ref{fig:NS_summary_plot}c). A notable trend for non-secular orbits is that a large majority produce a mean clockwise spin for Weeyo, which mainly results from the mutual orientations of the lobes at the periapsis, given their initial doubly synchronous state in the pre-phase circular orbit (Figure~\ref{fig:NS_summary_plot}d).

For the precession orbits, the mean collision angle exhibits a strong clustering between $85^\circ$ and $115^\circ$ across the entire investigated parameter space (Figure~\ref{fig:PR_summary_plot}a). No clear correlation is observed between the system's in-plane prolateness or bulk density and the resulting mean collision angle. Gas-to-gravitational torque ratios consistently range between $0.1 \%$ and $0.32 \%$. As established previously, an increase in bulk density results in a lower ratio regardless of variations in lobe shape (Figure~\ref{fig:PR_summary_plot}b). The distribution of short-axis spin ratios for Wenu and Weeyo reveals distinct dynamical behaviors for each lobe (Figure~\ref{fig:PR_summary_plot}c, d). While there is no clear linear trend between these ratios and the system's geometry or density, a notable difference in clustering is observed. Wenu's short-axis spin ratio remains tightly clustered around a baseline of $0.3$ to $0.4$ while Weeyo's short-axis spin ratio exhibits a much wider dispersion across the parameter space (Figure~\ref{fig:PR_summary_plot}d).

Analysis of the minimum-separation orbits reveals that the resulting collision angles show no correlation with the shapes and bulk densities of the lobes (Figure~\ref{fig:MS_summary_plot}a). These orbits produce mean collision angles concentrated within a narrow band between $87^\circ$ and $93^\circ$ across all simulated cases. This distribution is significantly higher than Arrokoth’s observed alignment of $6^\circ$ \citep{Spencer2020ArrokothGeology}. Across all simulations, gravitational torque overwhelmingly dominates gas torque (Figure~\ref{fig:MS_summary_plot}b). This dominance is particularly pronounced because its closer distance than other orbital cases maximizes mutual gravitational torque while near-planar motion minimizes the incidence angle and resulting gas torque. Furthermore, the gas drag-to-gravity ratio exhibits the characteristic inverse relationship with bulk density established in previous orbital configurations. Despite these variations, both Weeyo and Wenu effectively maintain pure short-axis spin throughout the pre-collision phase (Figure~\ref{fig:MS_summary_plot}c, d).

\begin{figure}[H]
    \centering
    \includegraphics[width= 0.6\linewidth]{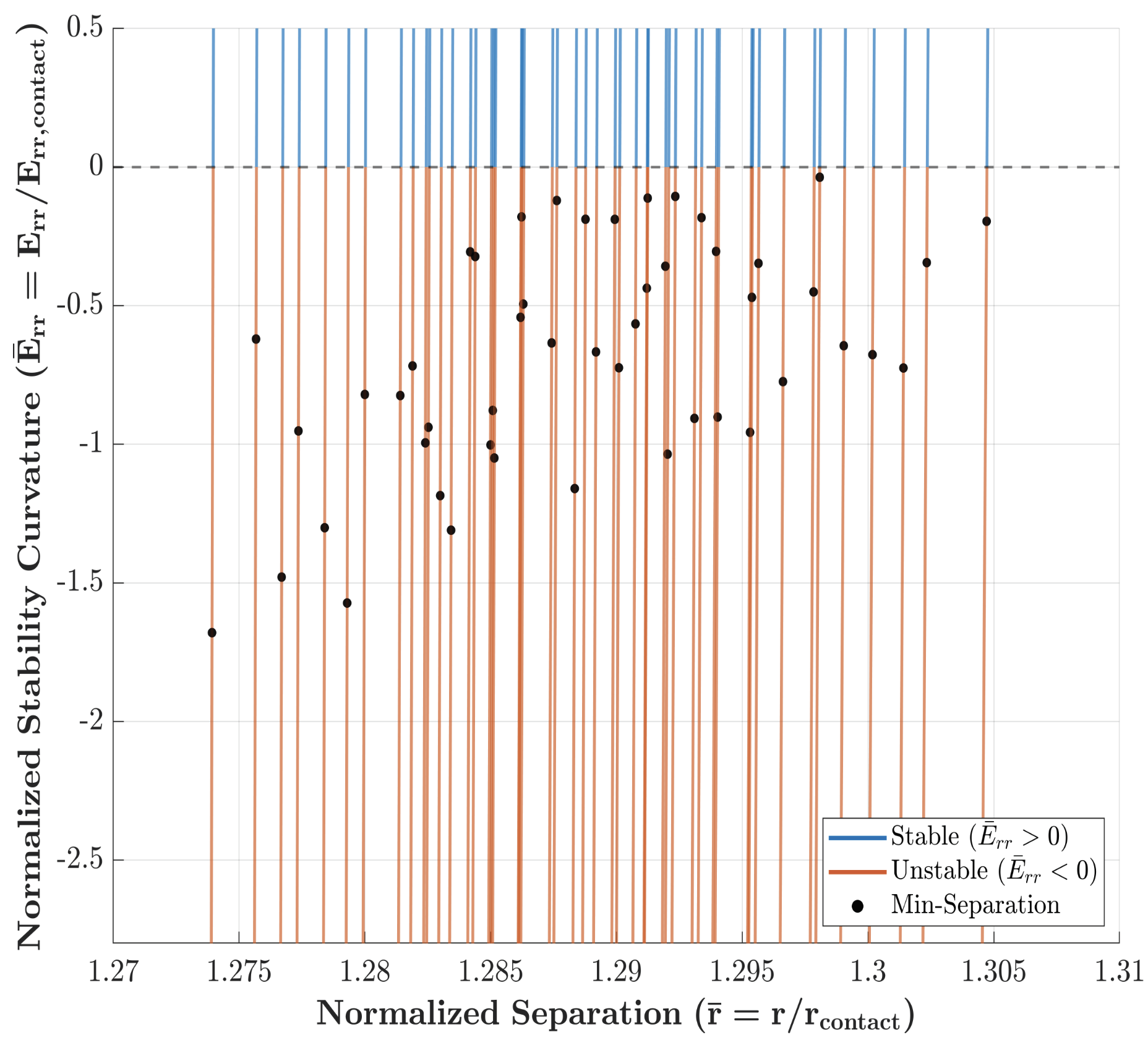}
    \caption{Normalized stability curvature $\bar{E}_{rr}$ as a function of the normalized separation $\bar{r}$. The normalization factor $r_{contact}$ is the sum of the long semi-axes for each specific lobe shape combination, ranging between 17.94 and 19.03~km. The solid curve denotes the transition from stable (blue) to unstable (orange) energetic stability {\citep{scheeres2009stability}}. The black dots represent the initial separations for the min-separation orbits, which are initialized in an unstable regime.}
    \label{fig:stability_plot}
\end{figure}

The accelerated departure from relative equilibrium observed near contact results from the system's inherent energetic instability. A relative equilibrium can only be maintained when the stability curvature is positive (${E}_{rr} > 0$) \citep{scheeres2009stability}. The critical stability separation at which ${E}_{rr}$ transitions from positive to negative depends on the lobes' individual shapes, rather than their masses. Figure~{\ref{fig:stability_plot}} evaluates this condition by plotting the normalized stability curvature $\bar{E}_{rr} = E_{rr} / E_{rr,contact}$  against the normalized separation $\bar{r} = r / r_{contact}$. $r_{contact}$ is the sum of the lobes' longest semi-axes and lies between 17.94--19.03~km across our investigated shape combinations, while $E_{rr,contact}$ is the value of the stability curvature at this contact distance. As illustrated, all minimum-separation orbits (represented by black dots) are initialized just beneath their respective critical stability separations, such that $\bar{E}_{rr} < 0$.

\begin{figure}[h!] 
\centering
\includegraphics[width= 0.8\linewidth]{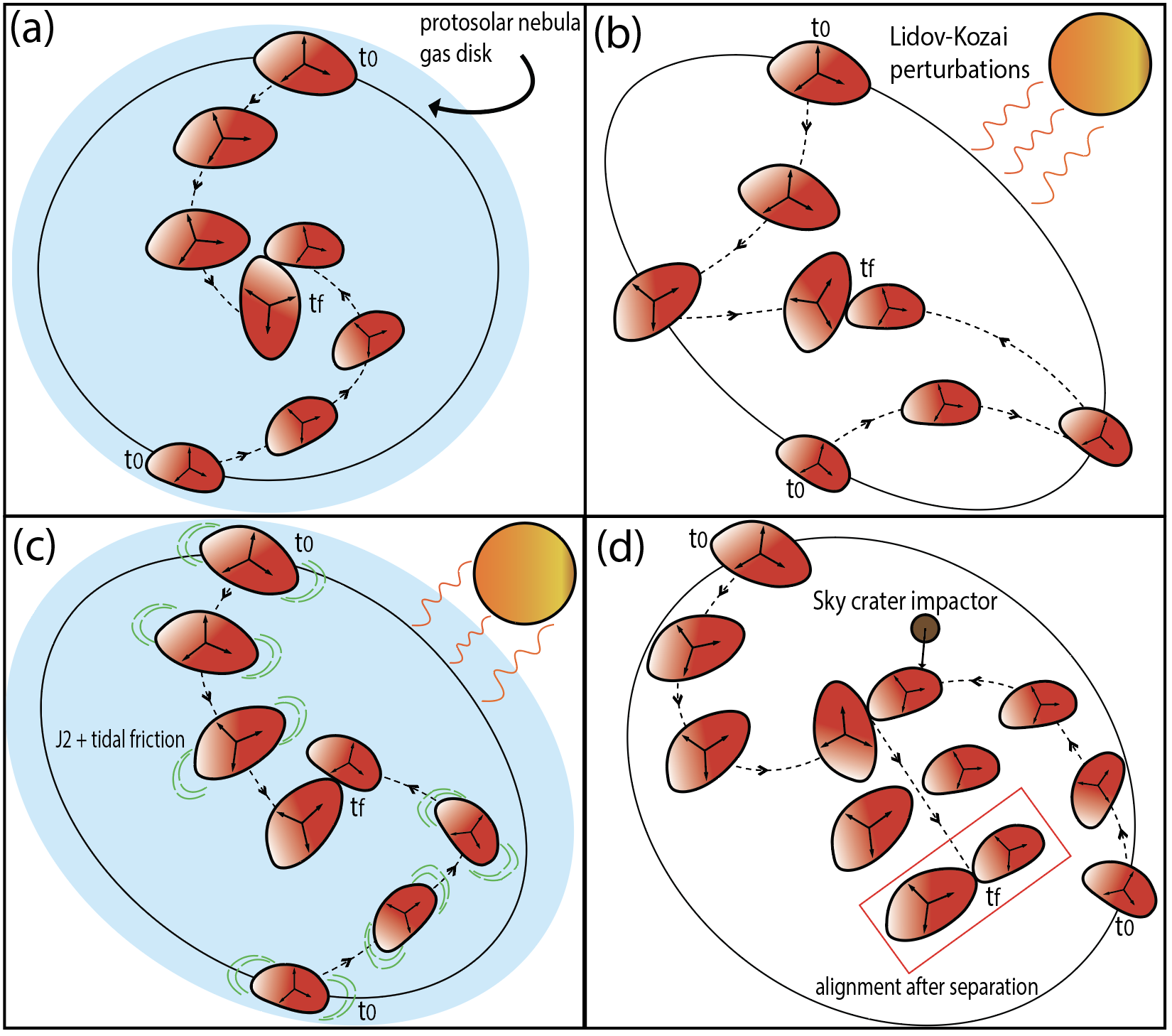}
\caption{Soft merger scenarios leading to Arrokoth's final principal axes alignment. 
Panel (a) shows that gas drag from the protosolar nebula gradually shrinks the binary orbit between Weeyo and Wenu, ultimately causing a merger \citep{McKinnon2020Arrokoth}. However, this mechanism alone is insufficient to stabilize the proximity behavior of the lobes and align their principal axes upon merging. Panel (b) illustrates that non-secular LK perturbations (shown in orange) trigger large oscillations in orbital eccentricity and separation \citep{Grishin2020WideBinary}. These perturbations can drive a collision within $\sim$1,000 years, but do not resolve the lobes’ unstable mutual dynamics. Panel (c) depicts a more comprehensive model that combines gas drag, $J_2$-induced quadrupole effects, LK perturbations, and tidal dissipation also leads to Weeyo and Wenu merging \citep{Lyra2021}. Despite the increased complexity and frequency of interaction between these forces, they cannot align the lobes during the merger. Panel (d) describes a hypothetical scenario that the Sky impactor reconfigured the initially misaligned merged configuration of Weeyo and Wenu \citep{Hirabayashi2020MarylandMU69, Kim2024CohesiveArrokoth,McKinnon2022SnowCrash}. This event caused a temporary separation of the lobes, ultimately resulting in the aligned configuration observed in Arrokoth today.
}
\label{fig:Arrokoth_LK}
\end{figure}

Expanding on these results, the previously theorized formation mechanisms fail to suppress this instability and produce a misaligned configuration between Weeyo and Wenu upon a one-time merger(Figures \ref{fig:Arrokoth_LK}a through \ref{fig:Arrokoth_LK}c). The resulting misalignment between Weeyo and Wenu upon a single merger is consistent with other observed contact binaries. For example, 25143 Itokawa exhibits a distinct misalignment between the principal axes of its lobes ($\sim$$24^\circ$ around the $\hat {\bfm z}$-axis and $\sim$$26^\circ$ around the $\hat {\bfm y}$-axis) \citep{Demura2006ItokawaShape}. Sub-catastrophic impact simulations also suggest complex shape configurations due to a soft merger \citep{Jutzi2019SubCatastrophicAsteroids,Jutzi2015CometaryAccretion}, casting an issue on the feasibility of Arrokoth's principal axis alignment after a single merger.

Building on this conclusion, we propose that a secondary event may have reconfigured an originally misaligned configuration. For the Arrokoth formation scenario, we theorize that the Sky-forming impactor released sufficient energy to temporarily disturb the lobes at the neck, ultimately resulting in their realignment into the observed configuration \citep{Hirabayashi2020MarylandMU69, Kim2024CohesiveArrokoth,McKinnon2022SnowCrash}. First, Weeyo and Wenu experience a soft merger based on the proposed orbital mechanisms, which does not offer the perfect principal axes alignment that Arrokoth exhibits (Figure \ref{fig:Arrokoth_LK}a through c). The subsequent Sky-forming impact disturbs the structural configuration \citep{Hirabayashi2020MarylandMU69, Kim2024CohesiveArrokoth}, potentially causing mechanical failure near the object's neck. While we suggest this impact as a plausible candidate for reconfiguration due to its established dynamics, we do not propose it as the sole possible mechanism. 
Calculations using the $\pi$-scaling law indicate that the impactor delivered a linear momentum of $2.4\text{--}4.0 \times 10^{13}$ kg m s$^{-1}$, yielding a kinetic energy transfer of approximately $1\text{--}3 \times 10^{12} \text{ J}$ to the small lobe \citep{Kim2024CohesiveArrokoth}. To survive this impact without structural failure, Arrokoth's cohesive strength would have had to exceed the typical 1~kPa cohesive strength of comets and asteroids by 20 times \citep{Kim2024CohesiveArrokoth}. Given this massive discrepancy, structural reconfiguration of Arrokoth is highly likely. Although some studies suggest the Sky-forming impact might not offer a strong structural disturbance \citep{McKinnon2022SnowCrash}, the level of stress disturbance remains similar across investigations.If the Sky-forming impact is indeed a trigger to break the neck, the resulting configuration minimizes the total energy under its angular momentum \citep{Hirabayashi2020MarylandMU69,Scheeres2007RotationalFission}, where the two lobes align along their principal axes (Figure \ref{fig:Arrokoth_LK}d).

\section{Conclusion} 
When simulating Weeyo and Wenu’s mutual dynamics using finite element modeling for the full two-body problem, it is demonstrated that their principal axes cannot align upon a single soft merger. The resulting misalignment between Weeyo and Wenu contradicts the aligned configuration between them presently observed in Arrokoth. The findings presented in this study also demonstrate that previously proposed formation mechanisms (e.g., LK perturbations, protosolar nebula gas drag) are unable to resolve the unstable rotational dynamics between Arrokoth’s lobes as they approach proximity. Consequently, the feasibility of Arrokoth forming via a single soft merger is strongly undermined. Our analysis suggests that the current alignment of Arrokoth’s lobes is likely the result of an external factor, such as the Sky impactor, which caused a structural disturbance at the neck of the object, leading to a reconfiguration into this alignment. 

\begin{acknowledgments}
This work was performed under 80NSSC24K0681 and with the support of the Partnership for an Advanced Computing Environment (PACE) at Georgia Tech.
\end{acknowledgments}

\clearpage  
%\bibliography{sample7_REVISED_Jan2026}{}
%\bibliographystyle{aasjournalv7}

%% the collaboration and author truncation commands are used. It has to


\end{document}